\documentclass[prl,aps, twocolumn,floatfix,longbibliography]{revtex4-2}
\usepackage[utf8]{inputenc}
\usepackage{natbib}
\usepackage{graphicx}
\usepackage{xcolor}
\usepackage[tbtags]{amsmath}
\usepackage{amssymb}
\usepackage{gensymb}
\usepackage{float}
\usepackage[normalem]{ulem}

\renewcommand{\vec}[1]{{\bf #1}}

\renewcommand{\phi}{\varphi}
\renewcommand{\epsilon}{\varepsilon}

\renewcommand{\dag}{\dagger}

\begin{document}

\title{Probing correlated states with plasmonic origami}

\author{Micha{\l} Papaj}
\thanks{These authors contributed equally to this work.}
\affiliation{Department of Physics, University of California, Berkeley, CA 94720, USA}
\author{Cyprian Lewandowski}
\thanks{These authors contributed equally to this work.}
\address{National High Magnetic Field Laboratory, Tallahassee, Florida, 32310, USA}
\address{Department of Physics, Florida State University, Tallahassee, Florida 32306, USA}

\begin{abstract}
Understanding the nature of strongly correlated states in flat-band materials (such as moir\'e heterostructures) is at the forefront of both experimental and theoretical pursuits. While magnetotransport, scanning probe, and optical techniques are often very successful in investigating the properties of the underlying order, the exact nature of the ground state often remains unknown. Here we propose to leverage strong light-matter coupling present in the flat-band systems to gain insight through dynamical dielectric response into the structure of the many-body ground state. We argue that due to the enlargement of the effective lattice of the system arising from correlations, conventional long-range plasmon becomes ``folded'' to yield a multiband plasmon spectrum. We detail several mechanisms through which the structure of the plasmon spectrum and that of the dynamical dielectric response is susceptible to the underlying order revealing valued insights such as the interaction-driven band gaps, spin-structure, and the order periodicity.
\end{abstract}

\maketitle

The moiré materials paradigm of combining 2D weakly-interacting materials to yield strongly-interacting systems is at the forefront of the current condensed matter research \cite{Balents2020,10.1038/s41563-020-00840-0,makSemiconductorMoireMaterials2022}. Moiré systems exhibit a wide range of phenomena ranging from unconventional superconductivity to various interaction-induced (correlated) resistive states \cite{cao1,cao2,10.1038/s41586-019-1695-0,Yankowitz1059, chenEvidenceGatetunableMott2019, chenSignaturesTunableSuperconductivity2019, ghiottoQuantumCriticalityTwisted2021, liContinuousMottTransition2021,wangCorrelatedElectronicPhases2020, tangSimulationHubbardModel2020, reganMottGeneralizedWigner2020, jinObservationMoireExcitons2019, seylerSignaturesMoiretrappedValley2019, tranEvidenceMoireExcitons2019, alexeevResonantlyHybridizedExcitons2019, liImagingTwodimensionalGeneralized2021}. Identification of the microscopic nature of the correlated states in the moiré systems is, however, difficult, as it relies on the interpretation of transport behavior or scanning-tunneling microscopy measurements \cite{10.1038/s41567-021-01438-2,doi:10.1126/science.abc2836,Ideue2021, Jiang2019,Kerelsky2019,Choi2019,PhysRevLett.129.117602,PhysRevLett.129.147001,2022arXiv220914322X,2022arXiv220912964Z,2022arXiv221101352Z}. To that end, despite the intense experimental and theoretical efforts, the exact flavor of the ground states is often not certain; thus, new tools to help identify them and complement existing results are in high demand. One such new approach can be based on the study of plasmons\cite{}, the collective charge excitations of interacting electron systems \cite{10.1038/nphoton.2012.262,10.1038/nphys2615,10.1038/nature01937, niFundamentalLimitsGraphene2018,10.1038/nature11254,10.1038/nature11253, zhaoEfficientFizeauDrag2021, dongFizeauDragGraphene2021}, as well as the overall properties of the system's dynamical dielectric response.

A defining characteristic of the moiré materials making their dielectric response distinct from that of conventional condensed matter systems, is the large effective lattice constant ($a_M \sim$ 10 nm)\cite{PhysRevLett.99.256802,Bistritzer12233, PhysRevB.95.075420}. The large unit cell size makes microscopic variations of the electric fields on the scale of the moiré period a significant effect, in contrast to the ordinary crystals with lattice constants $a \sim$ 0.1 nm necessitating consideration of local-field effects - i.e., treatment of screening effects accounting for the variation of the electric field within the unit cell \cite{wiserDielectricConstantLocal1963,adlerQuantumTheoryDielectric1962,PhysRevB.73.045112}. This phenomenon gives rise to a dynamical response function matrix $\epsilon_{GG'}(\vec{q},\omega)$, which depends not only on the frequency $\omega$ and momentum $\vec{q}$ inside the Brillouin zone (BZ), but is also labeled by the reciprocal lattice vectors $\mathbf{G}, \mathbf{G'}$ that relate Fourier components of the cell-periodic electric field (See Supplemental Materials for an overview). Crucially the dynamical response function matrix contains information about possible plasmon resonances, which are given by the solution of $\text{det}\, \epsilon_{GG'}(\vec{q},\omega)=0$ \cite{PhysRevB.7.2995,sturmHighfrequencyDielectricProperties1980,oliveiraHighfrequencyDielectricProperties1980}. Due to the matrix structure of the dielectric function, this characteristic equation can, in analogy to the free electron model, yield several branches of collective excitations. 

\begin{figure*}
    \centering
    \includegraphics[width=0.9\linewidth]{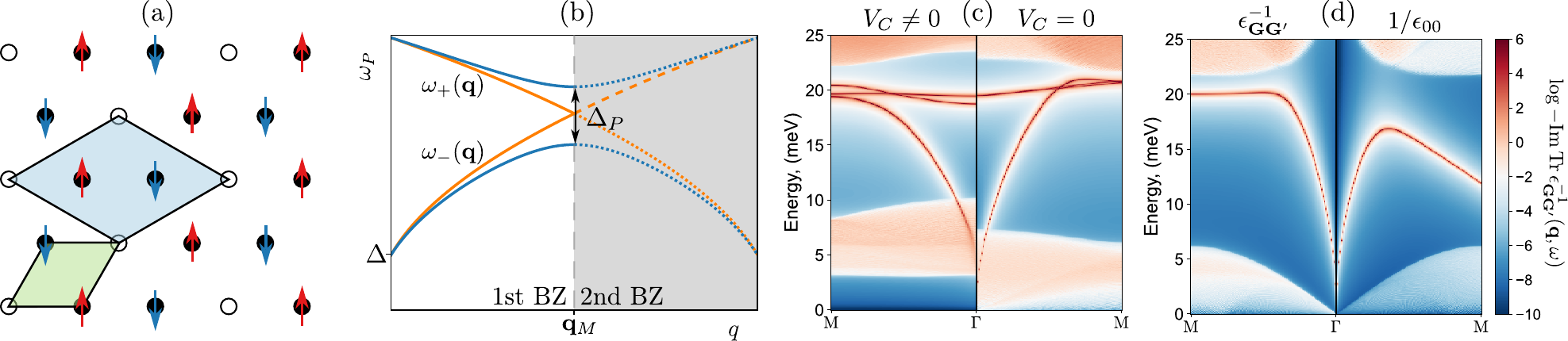}
    \caption{(a) With a correlated ground state, the moir\'e unit cell (green) expands to a larger effective cell (blue) that takes into account charge and spin ordering. An example is given for filling $\nu=2/3$ and antiferromagnetic state. (b) Schematic depiction of plasmon folding when crystal unit cell is extended. (c) The appearance of a correlated state introduces multiple plasmon branches and opens up gaps both between plasmon bands as well as in the particle-hole continuum. (d) In the absence of correlated state and unit cell enlargement, the inclusion of local field effects (left panel) doesn't introduce additional plasmon branches compared to when such effects are neglected (right panel).}
    \label{fig:fig_1}
\end{figure*}

In this work, we propose that the ``origami'' structure of the folded plasmon resonances, together with the overall behavior of the dynamical dielectric response, can allow for direct characterization of the microscopic nature of the correlated phases and their underlying ground states. While our findings are general and applicable to any moiré material, we focus on heterobilayer moiré transition metal dichalcogenides (TMD) \cite{wangCorrelatedElectronicPhases2020, tangSimulationHubbardModel2020, reganMottGeneralizedWigner2020, jinObservationMoireExcitons2019, seylerSignaturesMoiretrappedValley2019, tranEvidenceMoireExcitons2019, alexeevResonantlyHybridizedExcitons2019, liImagingTwodimensionalGeneralized2021}, where the difference in lattice constant of the materials in the two layers gives rise to the moiré pattern. Experimentally these systems exhibit interaction-induced insulating states at fractional and integer fillings consistent with generalized Wigner crystals pinned to the effective moiré lattice sites as demonstrated in Fig.~\ref{fig:fig_1}(a). Theoretically, as attributed to the low-energy band theory with a localized Wannier basis description, these insulating states are qualitatively well-described by a Hubbard model\cite{panQuantumPhaseDiagram2020, wuHubbardModelPhysics2018,PhysRevB.104.075150,2022arXiv220900664L}. Adopting this identification of the underlying competing ground states, we show how the different ground state flavors can result in drastically different dynamical responses, particularly in the plasmon spectrum, thus allowing for identifying such correlated states in experiments.

The formalism for treating local field effects has been introduced in the seminal works of Ref.~\citenum{wiserDielectricConstantLocal1963,adlerQuantumTheoryDielectric1962}. Specifically, when calculated with the random phase approximation (RPA), the dielectric function matrix takes the following form\cite{wiserDielectricConstantLocal1963,adlerQuantumTheoryDielectric1962,PhysRevB.73.045112}:
\begin{align}
    \epsilon_{\mathbf{G}\mathbf{G'}}(\omega, \mathbf{q}) =&\delta_{\mathbf{G}\mathbf{G'}}-T_{\vec{G}\vec{G'}}(\omega,\vec{q})\label{eq:epsGG_def} \\
      T_{\vec{G}\vec{G'}}(\omega,\vec{q})=&  V_{\mathbf{q} + \mathbf{G}} \sum_{n,m, \mathbf{k}} \frac{f_0(\epsilon_{n\mathbf{k}}) - f_0(\epsilon_{m\mathbf{k}+\mathbf{q}})}{\omega + i 0^+ + \epsilon_{n\mathbf{k}} - \epsilon_{m\mathbf{k} + \mathbf{q}}}
     \times   \nonumber\\
     &\times \eta_{\vec{q},\vec{G}}^{nm}(\vec{k})^* \eta_{\vec{q},\vec{G'}}^{nm}(\vec{k}) 
\end{align}
where the Fourier transform of the Coulomb potential is given by $ V_\mathbf{q} = \frac{2 \pi e^2}{|\mathbf{q}|}$, $f_0(\epsilon) = (e^{\beta(\epsilon-\mu)}+1)^{-1}$ (with $\mu$ the chemical potential and $\beta = 1/k_B T$ the inverse temperature). The state overlap $\eta_{\vec{q},\vec{G'}}^{nm}(\vec{k})$
\begin{align}
\label{eq:wavefunction_def}
   \eta_{\vec{q},\vec{G}}^{nm}(\vec{k}) = \frac{1}{\Omega} \int_{\Omega} d^2 \vec{r} ~u_{n \vec{k}}(\vec{r})^\dag e^{-i \vec{G}\cdot \vec{r}} u_{m \vec{k}+\vec{q}} (\vec{r})
\end{align}
is evaluated using $u_{n \vec{k}}(\vec{r})$, the cell-periodic part of the Bloch wavefunction $\psi_{n \vec{k}}(\vec{r}) = u_{n \vec{k}}(\vec{r}) e^{-i \vec{k} \cdot \vec{r}}$ for an eigenstate from band $n$ with a Brillouin-zone momentum $\vec{k}$ and energy $\epsilon_{n \vec{k}}$. The integral in Eq.~\eqref{eq:wavefunction_def} is taken over the unit cell with real-space area $\Omega$. Correspondingly, the displacement field due to the free charges is given by $\vec{D}_{\vec{G}}(\vec{q}) = \sum_{\vec{G'}} \epsilon_{\vec{G}\vec{G'}}(\omega,\vec{q}) \vec{E}_{\vec{G'}}(\vec{q})$, where $\vec{E}_{\vec{G'}}(\vec{q})$ are the Fourier components of the total electric field in the crystal. Plasmon, a sustained electric-field oscillation in the absence of free charges ($\vec{D}_{\vec{G}}(\vec{q}) = 0$) is given by the solution of a zero eigenvalue problem $\text{det}\, \epsilon_{\vec{G}\vec{G'}} = 0$. The relevant eigenvectors correspond to the real-space pattern of charge oscillations. Note that due to the finite size of the unit cell compared to the electromagnetic wave vector, the fields in free space do not map simply \cite{wiserDielectricConstantLocal1963,adlerQuantumTheoryDielectric1962} to the field in the crystal $\vec{E}_{\vec{G}}(\vec{q})$ due to the additional dependence on reciprocal lattice vectors (see Supplemental materials for further discussion). In a typical tight-binding model, where the cell-periodic part of the Bloch function becomes position independent ($u_{n \vec{k}}(\vec{r})=u_{n \vec{k}}$), the overlap of Eq.~\eqref{eq:wavefunction_def} vanishes for any $G\neq 0$ making the notion of local-field effects irrelevant (only $\epsilon_{00}(\omega,\vec{q})$ is non-zero in Eq.~\eqref{eq:epsGG_def}). This property is manifestly not true in continuum models of moiré materials where the cell-periodic part of the Bloch functions is position dependent, as we detail below.

To describe our platform of choice, the WSe$_2$/WS$_2$ heterobilayers with zero twist angle, we use the Bistrizer-MacDonald-type continuum model \cite{Bistritzer12233}. The valence moir\'e bands of this platform are comprised of WSe$_2$ hole pockets centered around K and K' points of the Brillouin zone (BZ). Owing to the spin-valley locking of TMDs \cite{xiaoCoupledSpinValley2012}, the K/K' valley degree of freedom can be identified with the electron spin, and the system is thus doubly degenerate. Within each valley, the Hamiltonian $H_0$ can be approximated by a parabolic band with periodic moir\'e potential $V_M(\mathbf{r})$ imposed on top of it \cite{wuHubbardModelPhysics2018, zhangMoirQuantumChemistry2020}:
 \begin{equation}
 \label{eq:ham_kinetic}
     H_0 = - \frac{\hbar^2 k^2}{2 m^*} + V_M(\mathbf{r}),
 \end{equation}
 where $m^* = 0.472\,m_e$ is the effective mass of the hole pocket of WSe$_2$. The moir\'e potential itself can be expressed in terms of its Fourier components corresponding to the first shell of reciprocal lattice vectors $\mathbf{b}_i$, with $\mathbf{b}_1 = 2\pi/a_M[2/\sqrt{3}, 0]$, and the remaining five vectors obtained by $\pi/3$ rotations. Here $a_M = a/\delta \approx 8.2$ nm is the moir\'e lattice constant where $a = 0.328$ nm and $\delta=4\%$ is the lattice constant of WSe$_2$ and lattice mismatch between WSe$_2$/WS$_2$ respectively\cite{wuHubbardModelPhysics2018, zhangMoirQuantumChemistry2020}. The moir\'e potential can thus be expanded as:
\begin{equation}
    V_M(\mathbf{r}) = \sum_i V_{\mathbf{b}_i} e^{i \mathbf{b}_i \cdot \mathbf{r}},
\end{equation}
where $V_{\mathbf{b}_1} = V_0 e^{i \psi}$, with $V_0 = 15\,\mathrm{meV}$ and $\psi=45\degree$ for WSe$_2$/WS$_2$ \cite{zhangMoirQuantumChemistry2020}, and the remaining coefficients can be obtained from symmetry as $V_\mathbf{b} = V_{R(2\pi/3, \mathbf{b})}$ and $V_\mathbf{b} = V^*_{-\mathbf{b}}$ (here $R(\theta,\mathbf{b})$ denotes counterclockwise rotation of vector $\mathbf{b}$ by $\theta$). 

The moir\'e potential breaks down the parabolic band from Eq.~\eqref{eq:ham_kinetic} into a series of mini-bands defined within the moiré Brillouin zone. The charge neutrality point of the entire structure lies within the bulk gap of WSe$_2$, and WS$_2$, thus the top-most doubly degenerate valence moir\'e band corresponds to the filling $-2 \le \nu \le 0$ as defined in the experiments \cite{xuCorrelatedInsulatingStates2020}. To study the fillings $0 \le \nu \le 2$, one can also use a similar model, which with an appropriate adjustment to the effective mass, describes moir\'e conduction bands formed by WS$_2$ electron pockets. In what follows when referring to a filling of $\nu$, we are considering a hole filling of valence bands $-\nu$ (i.e., the top valence band has $2-\nu$ electrons) or an electron filling $\nu$ of the bottom conduction band (i.e., the band has $\nu$ electrons in it). We also refer to these top valence or bottom conduction bands as the moir\'e ``flat band''.

To understand the origins of the correlated phases present in the TMD moiré structures, the flat band of the model can be considered within the framework of a Hubbard model on a triangular lattice \cite{panQuantumPhaseDiagram2020, wuHubbardModelPhysics2018,PhysRevB.104.075150,2022arXiv220900664L}. The correlated phases can then obtained using the Hartree-Fock mean-field treatment \cite{panBandTopologyHubbard2020, panQuantumPhaseDiagram2020,PhysRevB.104.075150} and depending on the strength of interaction and filling of the bands, the ground state can for example be ferromagnetic or antiferromagnetic with various patterns of in-plane or out-of-plane magnetic moments. In particular, we focus on the two key characteristics of the ground states that, as we will see, are relevant for the qualitative properties of the dielectric response: the enlargement of the unit cell due to the formation of a correlated phase, and what spin-structure the ground state has (c.f. Fig.\ref{fig:fig_1}a).  Specifically, building on the observation that the Hartree-Fock ground states have a generalized Wigner crystal character\cite{wuHubbardModelPhysics2018,panQuantumPhaseDiagram2020} with charge mainly localized around selected moir\'e potential minima, we use a Hartree-like electrostatic potential coming from the charges localized in the moir\'e potential minima and effective magnetic fields that are consistent with the orientation of the magnetic moments as shown in the self-consistent calculations. This approach, together with the enlargement of the unit cell due to the formation of the Wigner crystal and using experimentally-obtained estimates for the magnitudes of the correlated gaps \cite{xuCorrelatedInsulatingStates2020}, enables us to give an overview of the qualitative features observed in the dynamic dielectric response of the system that can help us to identify the nature of the correlated ground state.

Expressing the Hamiltonian $H_0$ in terms of reciprocal lattice vectors of the original moir\'e lattice $\mathbf{G}_0$ we have:
\begin{equation}
    H_0^{\mathbf{G}_0\mathbf{G}'_0}(\mathbf{k}) = -\frac{k^2}{2m^*}\delta_{\mathbf{G}_0\mathbf{G}'_0} + \sum_i V_{\mathbf{b}_i} \delta_{\mathbf{G}_0-\mathbf{G}'_0,\mathbf{b}_i}
\end{equation}
When a generalized Wigner crystal state appears, it is pinned to a fraction of the sites of the underlying moir\'e lattice, with the overall periodicity of that lattice different for each correlated ground state \cite{panBandTopologyHubbard2020, panQuantumPhaseDiagram2020,PhysRevB.104.075150}. This translates to a different set of reciprocal lattice vectors $\mathbf{G}$, which are shorter than moir\'e reciprocal lattice (i.e., moir\'e BZ becomes folded). In terms of the vectors $\mathbf{G}$, the Hartree (charge) potential takes the form
\begin{equation}
\label{eq:wigner_potential}
    H_C^{\mathbf{G}\mathbf{G'}}(\mathbf{k}) = V_C \sum_\mathbf{\tau} e^{-i (\mathbf{G}-\mathbf{G'})\cdot \mathbf{\tau}} V_{\mathbf{G}-\mathbf{G'}}\,
\end{equation}
where $V_C$ is a fitting parameter chosen to reproduce the experimental gap at a given filling as detailed in the concrete examples below, and $\mathbf{\tau}$ are the crystal basis vectors within the enlarged unit cell that specify the centers of charge distribution for a given filling. Similarly, the effective potential that enforces the spin texture predicted by Hartree-Fock calculations\cite{panBandTopologyHubbard2020, panQuantumPhaseDiagram2020} we take as given by:
\begin{equation}
\label{eq:spin_potential}
    H_M^{\mathbf{G}\mathbf{G}'}(\mathbf{k}) = V_B \sum_\mathbf{\tau} e^{-i (\mathbf{G}-\mathbf{G}')\cdot \mathbf{\tau}}  \mathbf{B}(\mathbf{G}-\mathbf{G}', \mathbf{\tau})\cdot \mathbf{\sigma}
\end{equation}
Here $V_B$ is also a fitting parameter, $\mathbf{B}(\mathbf{G}-\mathbf{G}', \mathbf{\tau})$ is a vector that determines the orientation of the magnetic field at a given crystal basis site $\mathbf{\tau}$, and $\mathbf{\sigma}$ is a vector of Pauli matrices that describes the spin degree of freedom.

Due to the folding of the moir\'e BZ, new electron bands form that originate from the moir\'e flat band. Once the effective charge (Eq.\eqref{eq:wigner_potential}) and spin (Eq.\eqref{eq:spin_potential}) potentials are included in the Hamiltonian gaps open between these folded bands. We, however, highlight that in each case, the model is only valid when the chemical potential is placed within the gap that corresponds to the filling fraction consistent with the ground state used. With the phenomenological models for correlated states described above, we can now calculate the energy loss function with the local field effects included. We perform the calculations for several distinct filling fractions $\nu = 1,\,2/3,\,1/2,\,1/3$ for different candidate ground states, which correspond to the most prominent generalized Wigner crystals as observed in the experiments \cite{xuCorrelatedInsulatingStates2020}. 

The central result of our work, the appearance of multiple plasmon resonances stemming from local-field effects and their strong dependence on the type of correlated order, is shown in Fig.~\ref{fig:fig_1}(b,c). This behavior can be understood by studying the structure of the dielectric function $\epsilon_{\vec{G} \vec{G'}}(\omega, \vec{q})$, which, in general,  has two types of entries: diagonal ($\vec{G}=\vec{G'}$) and off-diagonal ($\vec{G}\neq\vec{G'}$). The diagonal entries have the characteristic RPA structure of, $1-V_\vec{q} \Pi(\omega,\vec{q})\equiv 1-T_{\vec{G}\vec{G}}(\omega, \vec{q})$, each in principle yielding a collective excitation solution $\omega_{n}(\vec{q})$ when taken separately through $1=T_{\vec{G}\vec{G}}(\omega_{n}(\vec{q}), \vec{q})$ where the index $n$ labels each solution. The off-diagonal entries for frequencies outside of the particle-hole continua are of a fixed sign. As such, the diagonal entries give rise to the multiple folded (``origami'') plasmon branches, and the off-diagonal entries of the dielectric function matrix open gaps between them. 

To see this explicitly, let us focus on the behavior of $\epsilon_{\vec{G} \vec{G'}}(\omega, \vec{q})$ near the BZ edge at $\vec{q}_M=-\vec{G_1}/2$ (chosen to be the $M$ point of the hexagonal BZ) for the case of $\nu=2/3$ filling shown in Fig.~\ref{fig:fig_1}(c). In particular, we focus on a ground state which is a pure Wigner crystal ($V_C\neq 0$) without a spin-structure ($V_B=0$). Near $\vec{q}_M$, only the matrix entries corresponding to momenta $\vec{G}=0,\vec{G_1}$ contribute to leading order in the $\epsilon_{\vec{G} \vec{G'}}(\omega, \vec{q})$ due to the Coulomb prefactor $1/|\vec{q}+\vec{G}|$. The resulting dielectric function matrix has the structure
\begin{align}
\label{eq:dielectric_function_2by2}
\renewcommand*{\arraystretch}{2}
 \epsilon_{\vec{G} \vec{G'}}(\omega, \vec{q}) = \begin{bmatrix}
   1-T_{0 0}(\omega,\vec{q}) &
  -T_{0 \vec{G}_1}(\omega,\vec{q})\\
   -T_{\vec{G}_1 0}(\omega,\vec{q}) &
   1-T_{\vec{G}_1 \vec{G}_1}(\omega,\vec{q}) 
   \end{bmatrix}\,.
   \end{align}
For simplicity of the argument in evaluating the function $T_{\vec{G}\vec{G'}}(\omega,\vec{q})$ of Eq.~\eqref{eq:epsGG_def}, we can focus on the electron states closest in energy, i.e., bands separated by the correlated gap $\Delta$. To leading order in momentum $\vec{q}$ we can approximate the energy difference $|\epsilon_{n\mathbf{k}} - \epsilon_{m\mathbf{k} + \mathbf{q}}|\approx \Delta$ yielding
\begin{align}
\renewcommand*{\arraystretch}{2}
 \epsilon_{\vec{G} \vec{G'}}(\omega, \vec{q}) \approx \begin{bmatrix}
   1-\frac{A(\vec{q})}{\omega^2-\Delta^2} &
  -\frac{C(\vec{q})}{\omega^2-\Delta^2}\\
   -\frac{D(\vec{q})}{\omega^2-\Delta^2} &
   1-\frac{B(\vec{q})}{\omega^2-\Delta^2} 
   \end{bmatrix}\,.
   \end{align}
In arriving at this expression, we employed time-reversal symmetry of the system that requires $\omega \to -\omega$ (see Supplemental materials for careful derivation.) Resulting plasmon dispersion, (obtained from $\det \epsilon_{\vec{G} \vec{G'}}(\omega, \vec{q})=0$, see Fig.~\ref{fig:fig_1}(c)), is given by
\begin{equation}
    \omega_{\pm}^2(\vec{q}) = \Delta^2+\frac{1}{2} \left(A+B\pm \sqrt{(A-B)^2+4CD}\right)\,,
\end{equation}
where we suppress the dependence on momentum in $A, B, C$, and $D$ for clarity. We see that in the absence of the off-diagonal entries ($C=D=0$), there are two plasmon solutions, $\omega_-^2=\Delta^2+A(\vec{q})$ and $\omega_+^2=\Delta^2+B(\vec{q})$ (orange lines in Fig.~\ref{fig:fig_1}(b)). Off-diagonal entries hybridize the two solutions opening the gap in the plasmon spectrum at the boundary with the magnitude of the gap set by the Wigner potential. The other plasmon resonances seen in Fig.~\ref{fig:fig_1}(b)  originate similarly from the diagonal entries of $\epsilon_{\vec{G} \vec{G'}}(\omega, \vec{q})$ (see Supplemental Materials), however the energy ordering in which the individual resonances appear once off-diagonal entries are introduced is non-generic. We discuss both of these points in the following paragraphs.

The appearance of multiple plasmon branches is a robust feature expected simply due to the folding of the moir\'e BZ owing to the correlated order enlarging the unit cell, Fig.~\ref{fig:fig_1}(a). To see this explicitly, consider the bare Hamiltonian of Eq.~\eqref{eq:ham_kinetic} ($V_C=0$, $V_B=0$) but with an artificial enlargement of a unit cell to that of a unit cell of the $\nu=2/3$ Wigner crystal. The resulting plasmon resonance Fig.~\ref{fig:fig_1}(c) is folded at the new BZ edge, c.f. Fig.~\ref{fig:fig_1}(b), with a gap opening up in the plasmon dispersion when an actual correlated order appears, in addition to the opening of a gap $\Delta$ in the band structure which gaps out the conventional intraband plasmon as $\vec{q}\to0$.

\begin{figure*}[t]
    \centering
    \includegraphics[width=0.9\linewidth]{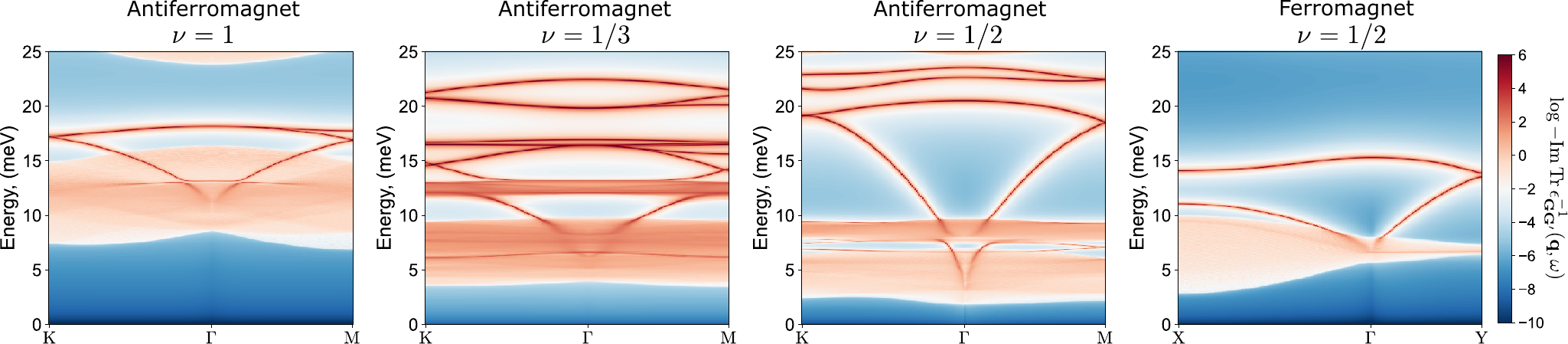}
    \caption{The trace of the electron loss function at $\nu = 1, 1/3, 1/2$ hole filling fractions. In each case, multiple plasmon branches are present due to local field effects. The enlargement of the moir\'e lattice unit cell in the correlated state determines the number of the plasmon branches. The size of the interacting gap determines the lower bound of interband particle-hole continua. In the case of $\nu = 1/2$, the two competing states are ferromagnetic and antiferromagnetic, and due to their different lattice periodicity, they have radically different plasmonic spectra.}
    \label{fig:fig_2}
\end{figure*}

We pause here to comment on the key claim of our paper, which attributes the appearance of the plasmon folding solely to the correlated order. The dielectric function matrix is a well-studied concept, both in first-principle calculations and early works on properties of nearly-free electron systems \cite{wiserDielectricConstantLocal1963,adlerQuantumTheoryDielectric1962,PhysRevB.73.045112}. In fact, Ref. \cite{oliveiraHighfrequencyDielectricProperties1980} first predicted the appearance of the folded plasmon resonance, albeit for a narrow momentum window (greatly limited by damping) in a nearly-free 3D electron gas. In the flat band systems, due to the mismatch between the kinetic and Coulomb energy scales, plasmon dispersion rises above the particle-hole continuum \cite{Lewandowski20869,PhysRevB.102.125403,PhysRevB.102.125408, PhysRevB.103.115431,PhysRevLett.125.066801}, circumventing the problem encountered by Ref.~\cite{oliveiraHighfrequencyDielectricProperties1980}. At first glance, therefore, a metallic moir\'e flat band system should exhibit such folded plasmon resonances. This, however, does not occur due to the subtle interplay of interband and intraband contributions to the polarization functions, as we explain below.

When local field effects are introduced in a metallic moir\'e flat band system, c.f, Fig.~\ref{fig:fig_1}(d), the plasmon dispersion flattens and is pushed to higher energies near the BZ, but a new plasmon branch does not appear. Authors of Ref. \cite{ceaCoulombInteractionPhonons2021} found similar behavior in the case of the twisted bilayer graphene continuum model, lacking any additional plasmon resonances. The origin of this behavior can be traced to the form of the dielectric response of the Eq. ~\eqref{eq:dielectric_function_2by2}. Schematically, the $\vec{G}_1\vec{G}_1$ entry can be found by extending the $00$ entry to the second BZ and then folding it back into the first BZ as depicted in Fig.~\ref{fig:fig_1}(b). Suppose the plasmon dispersion has a maximum away from the BZ boundary. In that case, the resulting folded branch (i.e., found only by keeping the diagonal entries of the dielectric matrix) will appear below the original plasmon branch. In turn, when the off-diagonal elements of the dielectric matrix are considered, the resulting gap between the original (now higher energy) branch and that of the folded (now lower energy) branch pushes the folded branch into the particle-hole continuum and the original branch upwards in energy, flattening it near the BZ boundary as shown in Fig. \ref{fig:fig_1}(d).
On the other hand, suppose the original plasmon branch has a maximum at the BZ edge, as it does in the schematic of Fig.\ref{fig:fig_1}(b). In that case, the folded plasmon will appear in the BZ above the original plasmon branch. The off-diagonal element of the dielectric matrix will then separate the two plasmons that are not pushed into the particle-hole continuum.

A nontrivial question therefore arises: What controls the overall shape of the plasmon dispersion? Or, more precisely: Where does its maximum in dispersion lie in relation to the BZ edge? Due to the interplay of two types of entries in the polarization function, as discussed in Ref.~\cite{Lewandowski20869}, those corresponding to transitions between the states separated by energies higher than the plasmon resonance (type I), and those corresponding to transitions between states separated by energies lower than the plasmon resonance (type II). The type II transitions give rise to the plasmon dispersion $\propto \sqrt{q}$, while the type I transitions act to renormalize the plasmon dispersion at large momenta comparable to the BZ scale. As a result of this interplay, in a generic metallic moir\'e flat band system, a maximum in the plasmon dispersion will occur away from the BZ boundary. When a correlated order appears, the maximum of the plasmon dispersion is pushed toward the BZ zone interface, and the plasmon becomes flattened (See Supplemental Materials for further discussion). This process, in conjunction with the reduction of the BZ size, gives rise to the ``origami''-like folding of plasmons, which are flattened at large momenta, and hence produce many weakly-dispersing plasmon resonances. This rich interplay of type I and type II transitions with local-field effects appears only in multiband continuum models and thus is missed by a preliminary tight-binding-based analysis as studied in Ref.~\cite{fahimniyaDipoleactiveCollectiveExcitations2020}, which  predicts folding of a plasmon branch in any two-site model - e.g., both metallic twisted bilayer graphene as well as a Mott insulator would manifest folded plasmon branches.

We can now analyze the different dynamical responses for the various filling fractions and candidate ground states in more detail; see Fig.~\ref{fig:fig_2}. Unlike the uncorrelated state, Fig.~\ref{fig:fig_1}(d), here, for any of the fractional fillings, we find several plasmon branches with the precise number set by the underlying ground state. More specifically, based on the arguments of the previous section, if only the diagonal entries were considered, we should see many plasmon branches $\epsilon_{\vec{G} \vec{G'}}(\omega, \vec{q})$. Here, however (despite considering more than 40 reciprocal lattice vectors in the presented results), we see significantly fewer resonances, i.e., the off-diagonal components of the dielectric function push other resonances into the particle-hole continua. Surprisingly, we also find that the number of plasmon resonances (degenerate along some high-symmetry directions) is precisely equal to the enlargement of the unit cell compared to the original moir\'e BZ, i.e., for $\nu=1$, we expect $3$,  for $\nu=1/3$ we expect $9$, for $\nu=1/2$ (AFM) we expect $4$ and for $\nu=1/2$ (FM) we expect $2$. We postulate this to be a key signature that enables the detection of the underlying lattice enlargement.

\begin{figure*}[t]
    \centering
    \includegraphics[width=0.9\linewidth]{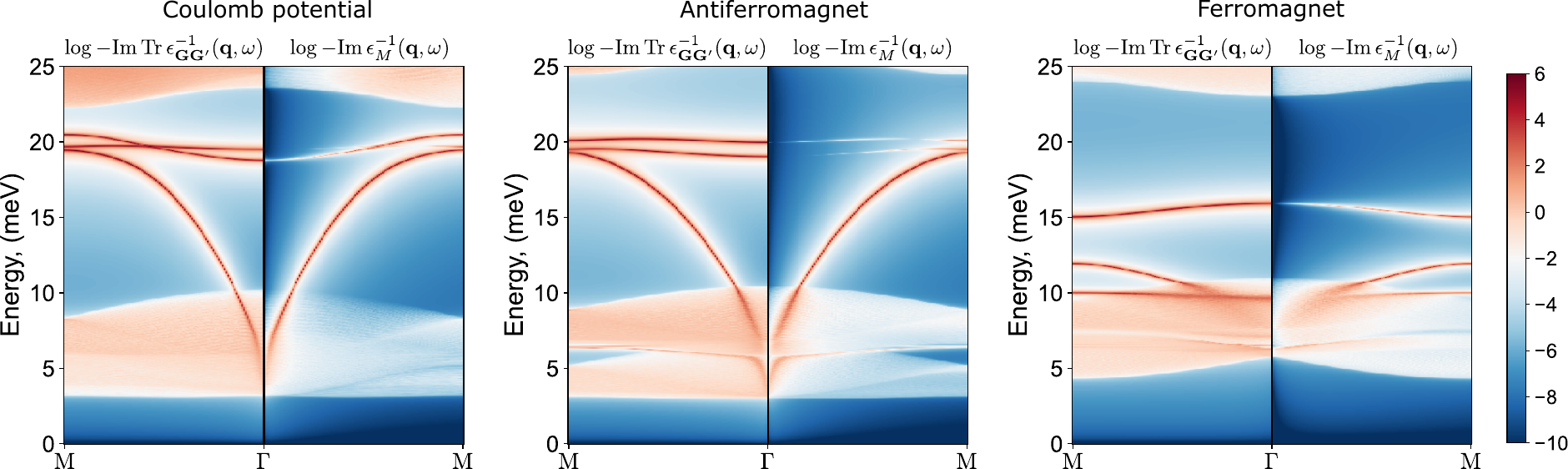}
    \caption{The comparison between the trace of the energy loss function and the inverse macroscopic dielectric function\cite{wiserDielectricConstantLocal1963,adlerQuantumTheoryDielectric1962,PhysRevB.73.045112} for three different ground states at filling $\nu = 2/3$. Even though, in each case, the unit cell of the correlated state is the same (triple the size of the moir\'e unit cell), the plasmon spectra and particle-hole continua are distinct. While the trace of the loss functions shows all of the possible collective excitations of the system, the macroscopic dielectric function shows the optical absorption spectrum.}
    \label{fig:fig_3}
\end{figure*}

The details of the plasmon dispersion behavior and the overall dynamical response (specifically, the particle-hole continuum) can yield further insights into the nature of the underlying ground state. The interband continuum reveals the presence and magnitude of the correlated gaps as it extends down to the lowest energies in the metallic case. Similarly, the plasmon dispersion in the metallic case stretches down to zero frequency as $\mathbf{q} \rightarrow 0$. In contrast, when interactions open the gap, the plasmon also rises to higher energies, even for the smallest momenta, as discussed in the context of Fig.~\ref{fig:fig_1}(b,c). As the moir\'e materials allow for continuous tuning of the filling fraction, this opening of the gap could be directly observed when the plasmon spectrum is measured. 

Our method can also help distinguish between different types of correlated ground states that can appear at the same filling. For example, at $\nu = 1/2$, the generalized Wigner crystal forms stripes, separated by two moir\'e lattice constants \cite{panQuantumPhaseDiagram2020}. However, such stripes may have either ferromagnetic (FM) or antiferromagnetic (AFM) chain configurations. While in the FM state, the lattice period becomes doubled only in one direction, the AFM state has a doubling of both lattice vectors. This corresponds to two and four times smaller Brillouin zones for FM and AFM, respectively. This, as shown in Fig.~\ref{fig:fig_2}(c,d), has dramatic consequences for the dielectric response. Not only the plasmon energies are entirely different, but even the number of plasmon branches changes between the two states, as mentioned previously. This enables distinguishing the type of ground state present in the system.

However, even if the unit cell is the same for two magnetic structures at the same filling, it is still possible to differentiate them based on the plasmon spectrum. This is demonstrated in Fig.~\ref{fig:fig_3}, where we show the results at $\nu = 2/3$ for three different correlated states. They each have the same effective unit cells but have either purely electrostatic potential, an antiferromagnetic state, or a ferromagnetic state, and in each case, the correlated gap is set to 2.9 meV. Apart from the differences in plasmon dispersion, there is another striking contrast between the FM and AFM states: owing to the spin-polarization in the FM state, some excitation processes are forbidden due to the vanishing matrix elements of band overlap. This leads to an effective gap of 3.9 meV as determined by the boundaries of the particle-hole continuum. Such a gap would thus be different from the one observed in STM or activated transport experiments, where the matrix elements do not play a role. This adds an additional feature through which the dielectric response can identify these ground states.

Throughout the manuscript, we plotted the imaginary part of the trace of the inverse dielectric function, $\mathrm{Tr}\, \epsilon_{\vec{G}\vec{G'}}^{-1}(\vec{q},\omega) \propto 1/\det \epsilon_{\vec{G}\vec{G'}}(\vec{q},\omega)$, which carries information about all plasmon resonances ($\det \epsilon_{\vec{G}\vec{G'}}(\vec{q},\omega)=0$). However, an experimentally relevant question is whether all of these modes can be excited, particularly those plasmon resonances corresponding to the long-wavelength limit $\vec{q}\to 0$. A detailed answer to this question depends on the specific experimental method used, as different techniques may excite different patterns of charge oscillations (e.g., dipole, etc.). For conventional optical probes ($\vec{q}\to 0$), the relevant quantity is the loss function defined \cite{Martin2016-dc} as $\mathrm{log} -\epsilon_{M}^{-1}(\vec{q},\omega)$ where $\epsilon_M(\vec{q},\omega)$ is the macroscopic dielectric function \cite{wiserDielectricConstantLocal1963,adlerQuantumTheoryDielectric1962,PhysRevB.73.045112}
\begin{equation}
\frac{1}{\epsilon_M(\vec{q},\omega)} = (\epsilon_{\vec{G}\vec{G'}}^{-1}(\vec{q},\omega))_{00}\,,
\end{equation}
that is the $\vec{G}=\vec{G'}=0$ entry of the inverse of the dielectric function from Eq.~\eqref{eq:epsGG_def}. This loss function describes the probability of exciting plane wave longitudinal charge oscillations. In Fig.~\ref{fig:fig_3}, we compare the spectrum of the plasmon excitations to the plasmons that can be excited by conventional optical probes.  We see that while some of the folded plasmons remain in the optical excitation spectrum with varying intensity, some of the resonances are not active and likely require experimental probes that can excite charge oscillations where individual Fourier wavevectors $E_{\vec{G}}$ have an internal structure (are not plane wave-like). We leave detailed consideration of these effects to future work. 

In summary, we have shown that the dynamical dielectric response, particularly the plasmon dispersion, can be a valuable tool in identifying correlated phases of moir\'e materials. The advent of tunable on-chip THz radiation sources presents an opportunity to clarify the nature of the rich phase diagram of correlated states in flat-band materials. Moreover, the results we show here only constitute the first step in investigating the phenomenon of plasmon origami-like folding. One possible future direction is the investigation of real space patterns of charge density oscillations, which could be observed using scanning near-field optical microscopy\cite{10.1038/nature11254, 10.1038/nature11253, PhysRevLett.119.247402, hesp2019collective}.

While the folding of the plasmon spectrum is a robust phenomenon stemming from the physics of BZ folding, the overall dispersion is sensitive to the details of the correlated order. In particular, in our analysis, we assume a simplified framework wherein the structure of the correlated order has no dependence on BZ momentum $\vec{q}$. This is an oversimplification of the analysis, which a calculation that involves Hartree-Fock treatment could remedy. Such a calculation could also treat RPA screening with Hartree-Fock self-consistently\cite{PhysRev.124.287,PhysRev.127.1391} to conserve the particle number and imbue the plasmons with an additional momentum and/or spin dependence. In particular, moir\'e systems can in principle exhibit other collective modes (e.g. spin-waves), which can similarly be obtained through the framework of RPA analysis. Thus consideration of ``local field effects'' in the context of such excitations may prove a fruitful research direction to follow, potentially yielding alternative probes into the microscopic nature of the correlated states in the moir\'e materials.

\begin{acknowledgments}
M.P. was supported by the Quantum Science Center (QSC), a National Quantum Information Science Research Center of the U.S. Department of Energy (DOE).  M.P. received additional fellowship support from the Emergent Phenomena in Quantum Systems program of the Gordon and Betty Moore Foundation. C.L. was supported by start-up funds from Florida State University and the National High Magnetic Field Laboratory. The National High Magnetic Field Laboratory is supported by the National Science Foundation through NSF/DMR-1644779 and the state of Florida. 
\end{acknowledgments}

\bibliography{references}

\pagebreak
\widetext

\setcounter{equation}{0}
\renewcommand{\theequation}{S\arabic{equation}}
\renewcommand{\thefigure}{S\arabic{figure}}

\end{document}